\documentclass{article}
\usepackage[utf8]{inputenc}
\pdfoutput=1
\usepackage{subfigure}
\usepackage{amsmath}
\usepackage{amssymb}
\usepackage{calrsfs}
\usepackage{dsfont}
\usepackage{bbm}
\usepackage[utf8]{inputenc}
\usepackage{graphicx}
\usepackage{float}
\graphicspath{{Images/}}
\DeclareMathAlphabet{\pazocal}{OMS}{zplm}{m}{n}
\usepackage{enumerate}
\usepackage{mathtools}
\usepackage[multiple]{footmisc}
\usepackage{bigfoot}
\usepackage{authblk}
\title{Brownian Motion \& The Stochastic Behaviour of Stocks}
\date{September 2021}
\author[1]{Yorgos Protonotarios}
\author[2]{Pantelis Tassopoulos}
\affil[1]{Department of Mathematics, University College London}
\affil[2]{Department of Mathematics, Imperial College London}

\begin{document}

\maketitle

\section{Introduction}
\subsection{Robert Brown}
In 1827, a botanist by the name of Robert Brown was examining the motion of grains of pollen suspended under water from a species of plants. Brown observed the motion of the particles  ejected from these pollen grains which followed a seemingly "jittery" motion; this was the first ever recorded case of Brownian motion, named after Robert Brown. The idea behind this type of motion being that the trajectory follows a completely random and "unpredictable" path. Since then, the concept of an unpredictable and random trajectory has been utilized in numerous fields including financial mathematics in the modeling of stock behaviour \cite{brown1828xxvii}.\\ 

\subsection{Louis Bachelier}
Several decades later, in 1900, Louis Bachelier built the foundation of mathematical finance by integrating Brownian motion with the fluctuation in the price of a stock. He postulated that two ideas should be considered when exploring the future value of an asset. First, how a collection of \emph{anterior} (past)  events influence the asset and second, how the probability of unknown future events could affect it \cite{bachelier1900theorie}. For example, the C.E.O of Apple just got replaced yesterday; that is an \emph{anterior} event that could influence the future price of an Apple stock. On the other hand, if a power outage occurs tomorrow in one of Apple's factories and halts production, that is an unknown future event that might affect the price of the stock and falls into the second category. It is said that the fluctuation in the price of a stock attributed to the latter follows Brownian motion as it is seemingly unpredictable and random. Bachelier only focused on the second idea since the \emph{anterior} events were not meaningful because Bachelier defined the mathematical expectation for an asset to rise or fall to be zero since the market constitutes of a pool of people that trade with opposite beliefs on the future value of an asset \cite{bachelier1900theorie}. For example, suppose a person is buying a call option for an asset, they hence believe that its price will rise. The person on the other side of the option believes it will fall otherwise they would not be selling the option. Hence, since they are both aware of past events that can influence the asset and have contradictory beliefs on how the it will perform, the mathematical expectation of gold is neither positive nor negative. The market is therefore said to be \emph{fair}. \cite{bachelier1900theorie}\\

\subsection{Norbert Wiener}
The above considerations by Bachelier were made concrete and expanded upon within a physics framework by American mathematician Norbert Wiener in his seminal work on 'differential spaces' \cite{wiener1923differential}. This was justified by arguments pertaining to the motion of a particle suspended in a fluid, where the movement of the said particle depended on impulses by fluid particulates and the initial velocity of the particle, although the influence of the latter was deemed negligible by Einstein. For the sake of simplicity, one can assume that the particle is constrained in one dimension and so are the impulses. Due to the nature of the situation, Wiener, following Einstein assumed that the displacement of the particle between any two time instants had no bias in any direction and large movements relative to the time scale were unlikely. This prompted the use of the normal distribution as a way of describing this behaviour. A key assumption of the paper in the construction of Brownian motion is that these displacements are independent and normally distributed constitute the 'dimensions', or independent variable of the motion.

\section{Definition of Brownian Motion}
In probability theory, one usually considers three objects when setting up a probability space; namely, the sample space $\Omega$- the collection of outcomes of a random process, a sigma algebra $\mathcal F$- the set of all measurable events and the probability measure $\mathbb P$ that measures the probability of said events that obeys certain axioms as laid out by Kolmogorov \cite{kolmogorov2018foundations}.
Consider the probability space $(\Omega, \mathcal{F}, \mathbb{P})$.\\
\\A stochastic process defined on the probability space $(\Omega, \mathcal{F}, \mathbb{P})$ is a measurable function $X:[0,\infty) \times \Omega \rightarrow \mathbb{R}$. Such a process $\mathfrak{B}(t,\omega)$ is called a Brownian motion if it satisfies the following conditions \cite{2006}:
\begin{enumerate}[I]
\item $\mathbb{P}(\omega; \mathfrak{B}(0, t) = 0) = 1$
\item For any $0 \leq s <t$, the random variable $\mathfrak{B}(t) - \mathfrak{B}(s)$ is normally distributed with mean $0$ and variance $t-s$, i.e\\
\begin{center}
$\mathbb{P}(a \leq \mathfrak{B}(t) - \mathfrak{B}(s) \leq b) = \displaystyle \frac{1}{\sqrt{2\pi(t-s)}}\int_{a}^{b}e^{{\frac{-x^2}{2(t-s)}}}dx$
\end{center}
\item $\mathfrak{B}(t,\omega)$ has independent increments, i.e for any $0 \leq t_{1} \leq t_{2} \leq ... \leq t_{n}$, the random variables
\begin{center}
    $\mathfrak{B}(t_{1}), \mathfrak{B}(t_{2}) - \mathfrak{B}(t_{1}), ..., \mathfrak{B}(t_{n})-\mathfrak{B}(t_{n-1})$
\end{center}
are independent and identically distributed, as per II.
\item Almost all sample paths of $\mathfrak{B}(t, \omega)$ are continuous functions, i.e:
\begin{center}
$\mathbb{P}(\omega; \mathfrak{B}(.,\omega)$ is continuous) $= 1 $
\end{center}
\end{enumerate}
Note that the dependence on $\omega$ was dropped in property II and III for simplicity reasons.

\section{Applications to the Stock Market} \label{Application of SDE}
\subsection{Stochastic Differential Equations}
We now consider a stock that is tradable on a stock market, such as an S\&P 500 stock like Apple and model it by a stochastic process  $Y_t(\omega):[0,\infty) \times \Omega \rightarrow \mathbb{R}$ on the probability space $(\Omega, \mathcal{F}, \mathbb{P})$. \\
\vspace{-0.1 cm}
\begin{center}
So what factors drive the process $\mathcal Y_t$?\\
\end{center}
\vspace{0.1 cm}
Well, an attempt at answering the above is to view the stock price $\mathcal Y_t$ as a sum of a deterministic component and a stochastic component containing 'noise' meant to represent an underlying uncertainty. Thus, for a small increment in time from $t$ to $t + dt$, the change in $\log \mathcal Y_t$ i.e., $d\log \mathcal Y_t$ is given by\\
\begin{center}
$d \log \mathcal Y_t = \gamma(t)\cdot dt + \sigma(t, \mathcal Y_t) \cdot$ "noise"\\
\end{center}
\vspace{0.1 cm}
where $\gamma(t)$ is the growth rate of a stock which is deterministic and $\sigma(t, \mathcal Y) \cdot$ "noise" is the volatility of the stock and is the stochastic component. \cite{oksendal2003stochastic}. The noise part can be modeled by a Brownian motion $\mathfrak B_t$, following Bachelier and Wiener. Thus, the above equation can be recast in the following form \cite{fernholz2002stochastic} \\
\begin{equation} 
\label{eq: sde}
d\log \mathcal Y_t = \gamma(t)\cdot dt + \sigma(t, \mathcal Y_t)\cdot d\mathfrak{B}(t) 
\end{equation}
where $d\mathfrak B_t = \mathfrak B_{t+dt}-\mathfrak B_{t}$. Equation \eqref{eq: sde} is an example of a stochastic differential equation \cite{oksendal2003stochastic}. In order to make precise what we mean by \eqref{eq: sde}, we consider a discretisation of the problem and consider the interval $[0,t]$ where $t\geq 0$. We further impose a partition $t_0=0<t_1<t_2<...<t_{k-1}<t=t_n$ of the interval and read \eqref{eq: sde} to mean:
\begin{gather*}
     \log \mathcal Y_{t_k} - \log \mathcal Y_{t_{k-1}}
     \\ = \gamma(t_{k-1})\cdot (t_k-t_{k-1}) + \sigma(t_{k-1}, \mathcal Y_{t_{k-1}})\cdot (\mathfrak B_{t_k}-\mathfrak B_{t_{k-1}})
\end{gather*}
for $k$ ranging from $0$ to $n-1$. A summation of the index $k$ yields the process

\begin{gather*}
    \displaystyle I_n[\mathcal Y_t](.) = \log \mathcal Y_{t_n \equiv t} - \log \mathcal Y_{t_{0}}\\
    =\sum_{k=0}^{n-1}\gamma(t_{k-1})\cdot (t_k-t_{k-1}) + \sum_{k=0}^{n-1} \sigma(t_{k-1}, \mathcal Y_{t_{k-1}})\cdot (\mathfrak B_{t_k}-\mathfrak B_{t_{k-1}})
    \label{eq: discrete sde}
\end{gather*}
 \cite{oksendal2003stochastic}. Now, in a certain sense, we have 'integrated the process' and have obtained an expression for the process at some time $t$, given an initial time $t_0$. Indeed, two summations that appear are reminiscent of discrete approximations to Riemann-Stieltjes integrals. One is tempted to take a limit of such partitions $\pi_n$ with mesh $|\pi_n| \equiv \max(t_{i}-t_{i-1})\rightarrow 0$ and obtain the corresponding equation
 \begin{gather*}
     \displaystyle \log \mathcal Y_{t_n \equiv t} - \log \mathcal Y_{t_{0}} 
     = \int_{0}^{t}\gamma(t)dt + "\int_{0}^{t}\sigma(t, \mathcal Y_t)d\mathfrak B_t"
 \end{gather*}
 The aim of the next chapter is to show that for suitably well behaved $\gamma(t)(\omega)$ and $f(t,\omega)\equiv \sigma(t, \mathcal Y_t(\omega))$, such a limit exists in the space $L^{2}(\Omega)$, the space of all square integrable random variables. This intuition will be made precise and explored in the next section, as we explore the Itô Integral. 
\section{Preliminaries}
\subsection{$\sigma$-algebras}
Consider the space $(X, \mathcal{A})$. A $\sigma$-algebra $\mathcal{A}$ \cite{bremaud2020probability} is a set of subsets in $X$ satisfying the following conditions:
\begin{enumerate}[I]
\item $X, \emptyset \in \mathcal{A}$
\item if $A \in \mathcal{A}$ then $\overline{A} \in \mathcal{A}$
\item if $A_{j} \in \mathcal{A}$ for $j \in \mathbb{N}$ then $\displaystyle \bigcup_{j \in \mathbb{N}} A_{j} \in \mathcal{A}$\\
\end{enumerate}
A \emph{generator} $\mathcal{G}$ is a set of arbitrary subsets such that these subset generate a $\sigma$-algebra denoted by $\sigma(\mathcal{G})$ \cite{bremaud2020probability}. This $\sigma$-algebra is also known as the smallest or \emph{minimal} $\sigma$-algebra. It is defined as:
\begin{equation*}
\sigma(\mathcal{G}) := \bigcap_{\substack{\hspace{0.04 in} \mathcal{H}_k \hspace{0.04 in} \text{is a} \hspace{0.04 in} \sigma-\text{alg}\\\hspace{0.1 in} \mathcal{G} \subset \mathcal{H}_k}} \mathcal{H}_k
\end{equation*}
for an arbitrary index $k$.

\subsection{Filtrations \& $\mathcal{F}_{t}$ - adapted processes}
The family $\mathbb{F} := \{\mathcal{F}_{t}\}_{t \in \bf{T}}$ of $\sigma$-algebras of $\mathcal{F}$ defined on a measurable space $(\Omega, \mathcal{F})$ is called a filtration if for all $s,t \in \bf{T}$ such that $s \leq t$,
\begin{equation*}
\mathcal{F}_{s} \subseteq \mathcal{F}_{t}
\end{equation*}
\vspace{0.2 in}
where $\bf{T}$ is an index set in any of the following: $\mathbb{R}, \mathbb{R}_{+}, \mathbb{N}, \mathbb{Z}$. \cite{bremaud2020probability}\\
An $\mathcal{F}_{t}$-measurable random variable signifies that:
\begin{equation*}
Z^{-1}(\mathcal{B}(\mathbb{R})) \subset \mathcal{F}_{t}
\end{equation*}
which is a shorthand notation for:
\begin{equation*}
Z^{-1}(B) \in \mathcal{F}_{t},  \forall B \in \mathcal{B}(\mathbb{R})
\end{equation*}\\
Where $\mathcal{B}(\mathbb{R}) = \sigma({[a, \infty)})$ for some $a \in
\mathbb{R}$.\\ 
Define a function $h: [0, \infty) \times \Omega \rightarrow \mathbb R$ with a filtration $\mathbb{F} \subset \Omega$. The function is then called $\mathcal{F}_{t}$-adapted if it is $\mathcal{F}_{t}$-measurable $\forall t \geq 0$. \cite{bremaud2020probability}\\
\subsection{A.M - G.M Inequality \& Lebesgue Integral}
The A.M - G.M inequality states the following:
\begin{equation*}
\frac{a^{'}+b^{'}}{2} \geq \sqrt{a^{'}b^{'}}
\end{equation*}
for $a^{'}, b^{'} \in \mathbb{R^{+}}$. Now replace $a^{'} = a^2$ and $b^{'} = b^2$. The inequality then becomes:
\begin{equation*}
\frac{a^{2} + b^{2}}{2} \geq |ab|.
\end{equation*}
for $a,b \in \mathbb{R}$. Using the fact that the Lebesgue integral with respect to the probability measure is monotonic, we can apply it to the inequality as follows:
\begin{gather*}
\int a^{2} + b^{2} d\mathbb{P} \geq 2 \int |ab| d\mathbb{P}\\
\mathbb{E}[a^{2} + b^{2}] \geq 2\mathbb{E} [|ab|]
\end{gather*}
\subsection{$L^{p}$ Spaces}
Suppose $X: \Omega \rightarrow \mathbb{R}^{n}$ is a random variable and $p \in [1, \infty)$. The $L^{p}$ norm of $X$, $\|X\|_{L^{p}}$ is defined as \cite{oksendal2003stochastic} :
\begin{gather*}
\displaystyle \|X\|_{L^{p}} = \Bigg(\int_{\Omega} |X(\omega)|^{p}d\mathbb{P}(\omega)\Bigg)^{1/p} = \Big(\mathbb{E}(|X|^{p})\Big)^{1/p}
\end{gather*}

\section{Itô Integration}
A process $\phi: [0, \infty) \times \Omega \rightarrow \mathbb R$ is called \emph{elementary} if:
\begin{equation*}
 \displaystyle \phi(t, \omega) = \sum_{j=1}^{K-1} Z_{j}(\omega) \mathbbm{1}_{(t_{i}, t_{i+1}]}(t)\\
\end{equation*}
where $0 \leq t_{1} < t_{2} < ... < t_K < \infty$ and where $Z_{j}$ $(1 \leq i \leq K)$ is a complex square-integrable $\mathcal{F}_{t_{i}}$-measurable random variable. \cite{bremaud2020probability}
 \begin{equation*} \displaystyle \int_{S}^{T}\phi(t,\omega)d\mathfrak{B}_{t}= \sum_{j=1}^{K-1} Z_{j}(\omega)[\mathfrak{B}_{t_{j+1}}-\mathfrak{B}_{t_{j}}](\omega)
 \end{equation*}
In our case, a square integrable random process is defined as
\begin{equation}
\mathbb{E}  \Bigg[\displaystyle \int_{S}^{T} |\phi(t)|^{2} dt \Bigg] < \infty
\end{equation}
Define $\mathfrak{L}(S,T)$ to be the class of functions \cite{oksendal2003stochastic}:
\begin{equation}
f: [0, \infty) \times \Omega \rightarrow \mathbb R
\end{equation}
such that:
\begin{enumerate}[I]
\item $f(t, \omega)$ is $\mathcal{B}(\mathbb{R}^{+}) \times \mathcal{F}$ - measurable
\item $f(t, \omega)$ is  $\mathcal{F}_{t}$-adapted
\item $\mathbb{E}  \Bigg[\displaystyle \int_{S}^{T} f(t,\omega)^{2} dt \Bigg] < \infty$
\end{enumerate}
\subsection{Construction of the Itô Integral}
We will know construct the Itô integral. We will omit the proofs of the steps needed for such a construction for the purpose of brevity. In brief, the construction comprises of approximation lemmas for processes $g \in \mathfrak{L}(S,T)$ in terms of elementary functions, and then utilise the isometry property of the ito integral for elementary processes and the completeness of the metric space $L^{2}(\Omega)$ to define a limit and call it the Itô integral. \vspace{0.1in} \\
{\bf STEP I}\\
Let $g \in \mathfrak{L}(S,T)$ be bounded and continuous for all $\omega \in \Omega$. Then there exists elementary processes $\phi_{n} \in \mathfrak{L}(S,T)$ such that  \cite{oksendal2003stochastic}
\begin{equation*}
\mathbb{E} \Bigg[\displaystyle \int_{S}^{T} (g-\phi_{n})^2 dt \Bigg] \rightarrow 0 \text{, \space as \space } n \rightarrow \infty
\end{equation*}\\
{\bf STEP II}\\
Let $h \in \mathfrak{L}(S,T)$ be bounded. Then there exist bounded functions $g_{n} \in \mathfrak{L}(S,T)$ such that they are continuous for all $\omega$  and $n$ and \cite{oksendal2003stochastic}

\begin{equation*}
\mathbb{E} \Bigg[\displaystyle \int_{S}^{T} (h-g_{n})^2 dt \Bigg] \rightarrow 0 \text{, \space as \space} n \rightarrow \infty\\
\end{equation*}
{\bf STEP III}\\
Let $f \in \mathfrak{L}(S,T)$. Then there exists a sequence $\{h_{n}\} \subset \mathfrak{L}(S,T)$ such that $h_{n}$ is bounded for each $n$ and \cite{oksendal2003stochastic}
\begin{equation*}
\mathbb{E} \Bigg[\displaystyle \int_{S}^{T} (f-h_{n})^2 dt \Bigg] \rightarrow 0 \text{, \space as \space} n \rightarrow \infty\\
\end{equation*}
Now, using the above steps we show that for $f, \phi_{n} \in \mathfrak{L}(S,T)$:
\begin{equation*}
\mathbb{E} \Bigg[\displaystyle \int_{S}^{T} |f-\phi_{n}|^2 dt \Bigg] \rightarrow 0 \text{, \space as \space} n \rightarrow \infty\\
\end{equation*}
First, maintain the definition for $f,\phi_{n}, g$ and $ h$ as they were in steps I, II and III. We have:
\begin{gather*}
\mathbb{E} \Bigg[\displaystyle \int_{S}^{T} |f-\phi_{n}|^2 dt \Bigg]\\ = \mathbb{E} \Bigg[\displaystyle \int_{S}^{T} |(f-h_{k})+(h_{k}-g_{m})+(g_{m}-\phi_{n})|^2 dt \Bigg]\\
\leq 2\mathbb{E} \Bigg[\displaystyle \int_{S}^{T} |f-h_{k}|^2dt\Bigg] + 4\mathbb{E} \Bigg[\displaystyle \int_{S}^{T} |h_{k}-g_{m}|^2dt\Bigg]
+ 4\mathbb{E} \Bigg[\displaystyle \int_{S}^{T} |g_{m}-\phi_{n}|^2dt\Bigg]
\end{gather*}
By Step III, fix a $k$ large enough such that:
\begin{equation*}
\mathbb{E} \Bigg[\displaystyle \int_{S}^{T} |f-h_{k}|^2dt\Bigg] < \frac{\epsilon}{6}\\
\end{equation*}
By Step II, fix an $m$ large enough such that:
\begin{gather*}
\mathbb{E} \Bigg[\displaystyle \int_{S}^{T} |h_{k}-g_{m}|^2dt\Bigg] < \frac{\epsilon}{12}
\end{gather*}
By Step I, fix an $N$ large enough such that for all $n \geq N$:
\begin{gather*}
\mathbb{E} \Bigg[\displaystyle \int_{S}^{T} |g_{m}-\phi_{n}|^2dt\Bigg] < \frac{\epsilon}{12}
\end{gather*}
Hence, for all $n \geq N$, combining the above parts yields that 
\begin{gather*}
    \mathbb{E} \Bigg[\displaystyle \int_{S}^{T} |f-\phi_{n}|^2 dt \Bigg]<\epsilon
\end{gather*} 
We now will state without proof a key property of the Itô integral for elementary processes $\phi \in \mathfrak{L}(S,T)$, known as {\it \bf Itô Isometry} and states the following and can be found in \cite{oksendal2003stochastic}
\begin{gather*}
    \displaystyle E\left[\left(\int_{S}^{T} \phi(t)d\mathfrak B_t\right)^{2}\right] = E\left[\int_{S}^{T} \phi(t)^{2}dt\right]
\end{gather*}
Now, armed with the above results, we are going to show that the sequence $\left\{\displaystyle \int_{S}^{T}\phi_{n}(t,\omega)d\mathfrak B_t\right\}_{n \in \mathbb N}$ is Cauchy in $L^{2}(\Omega)$, where the $\phi_{n}(t,\omega)$ are elementary approximants of $f(t,\omega)$ in the sense that $E\left[\int_{S}^{T}(f-\phi_n)^{2}dt\right]\rightarrow0$ as $n\rightarrow\infty$. For the above to be Cauchy, one needs
\begin{gather*}
    \displaystyle \Bigg \| \int_{S}^{T}(\phi_{n}(t,\omega)-\phi_{m}(t,\omega))d\mathfrak B_t\Bigg\|_{L^{2}(\Omega)}^{2}
    = E \left[ \left(\int_{S}^{T}(\phi_{n}(t,\omega)-\phi_{m}(t,\omega))d\mathfrak B_t \right) ^{2}\right]\\
    =  E \left[\int_{S}^{T}\left(\phi_{n}(t,\omega)-\phi_{m}(t,\omega)\right)^{2}dt\right]\\
    \leq 2 E \left[\int_{S}^{T}\left(f(t,\omega)-\phi_{n}(t,\omega)\right)^{2}dt\right]+ 2E \left[\int_{S}^{T}\left(f(t,\omega)-\phi_{m}(t,\omega)\right)^{2}dt\right] \rightarrow 0
\end{gather*}
as $n,m \rightarrow \infty$ by the A.M-G.M. inequality, the approximation lemma and the isometry property of the Itô integral. The completeness of $L^{2}(\Omega)$ implies that there is a random variable $I[f]_{S}^{T} \in L^{2}(\Omega)$ such that 
\begin{gather*}
\displaystyle E\left[ \left(I[f]_{S}^{T} - \int_{S}^{T}\phi_{n}(t,\omega)d\mathfrak B_{t}\right)^{2}\right] \rightarrow 0
\end{gather*}
as $n \rightarrow \infty$. We can now define the Itô integral of $f(t,\omega)$ as 
\begin{gather*}
    \displaystyle \int_{S}^{T}f(t,\omega)d\mathfrak B_{t} := I[f]_{S}^{T} \stackrel{L^{2}(\Omega)}{=\joinrel=} \displaystyle \lim_{ n \rightarrow \infty} \int_{S}^{T}\phi_{n}(t,\omega)d\mathfrak B_{t}
\end{gather*}
It is also important to mention that the above limit does not depend on the choice of approximants. So suppose the approximants $\phi_n(t,\omega)$ and $\psi_n(t,\omega)$ converge in the $L^{2}(\Omega)$ sense to $I_1[f]_{S}^{T}$ and $I_2[f]_{S}^{T}$ respectively, then we have

\begin{gather*}
    \Bigg \| I_1[f]_{S}^{T}-I_2[f]_{S}^{T}\Bigg \|_{L^{2}(\Omega)}^{2}\\
    \leq \Bigg \|  \left( I_1[f]_{S}^{T} - \int_{S}^{T}\phi_{n}(t,\omega)d\mathfrak B_t \right) - \left( I_2[f]_{S}^{T}- \int_{S}^{T}\phi_{n}(t,\omega)d\mathfrak B_t \right) \Bigg\|_{L^{2}(\Omega)}^{2}\\
    \leq \Bigg \| I_1[f]_{S}^{T} - \int_{S}^{T}\phi_{n}(t,\omega)d\mathfrak B_t \Bigg\|_{L^{2}(\Omega)}^{2} + \Bigg \| I_2[f]_{S}^{T} - \int_{S}^{T}\phi_{n}(t,\omega)d\mathfrak B_t  \Bigg\|_{L^{2}(\Omega)}^{2}\\
    \leq \Bigg \| I_1[f]_{S}^{T} - \int_{S}^{T}\phi_{n}(t,\omega)d\mathfrak B_t \Bigg\|_{L^{2}(\Omega)}^{2} + \Bigg \| I_2[f]_{S}^{T} - \int_{S}^{T}\psi_{n}(t,\omega)d\mathfrak B_t  \Bigg\|_{L^{2}(\Omega)}^{2} + \\
   \Bigg \| \int_{S}^{T}\phi_{n}(t,\omega)d\mathfrak B_t - \int_{S}^{T}\psi_{n}(t,\omega)d\mathfrak B_t  \Bigg\|_{L^{2}(\Omega)}^{2} \rightarrow 0
\end{gather*}
as $n \rightarrow \infty$.  This is because 
\begin{gather*}
    \Bigg \| \int_{S}^{T}\phi_{n}(t,\omega)d\mathfrak B_t - \int_{S}^{T}\psi_{n}(t,\omega)d\mathfrak B_t  \Bigg\|_{L^{2}(\Omega)}^{2} = \mathbb{E} \left[\int_{S}^{T}(\phi_{n}-\psi_{n})^{2}(t,\omega)dt\right]\\
    \leq  2\mathbb{E} \left[\int_{S}^{T}(\phi_{n}-f)^{2}(t,\omega)dt\right] + 2\mathbb{E} \left[\int_{S}^{T}(\psi_{n}-f)^{2}(t,\omega)dt\right] \rightarrow 0
\end{gather*} by the fact that $\phi_n$ and $\psi_n$ are approximants to $f$ and the definition of the ito integral.
This shows that $ \Bigg \| I_1[f]_{S}^{T}-I_2[f]_{S}^{T}\Bigg \|_{L^{2}(\Omega)}^{2} = 0$ which means that $\displaystyle \mathbb{P} \Big(\Big\{\omega \in \Omega \Big| I_1[f]_{S}^{T}(\omega)=I_2[f]_{S}^{T}(\omega)\Big\}\Big) = 1$, that is that the limits are 'almost surely' equal, hence the same in $L^{2}(\Omega)$.

\section{Numerical Results} \label{numerical}

\begin{figure}[H]
\begin{subfigure}
  \centering
  \includegraphics[width = 0.55\textwidth]{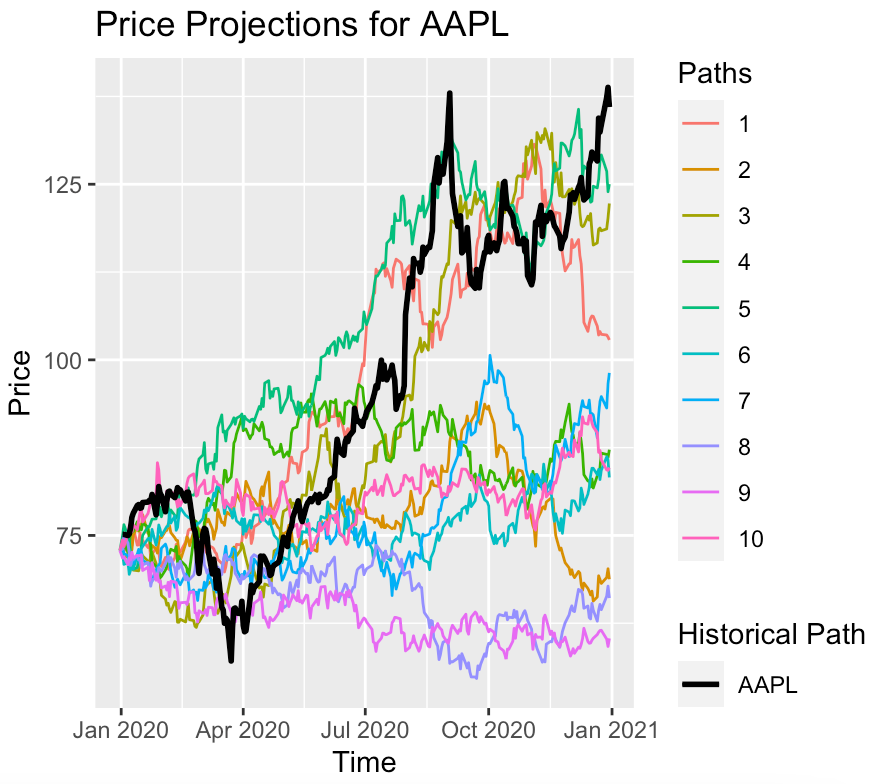}
  \label{fig:AAPL coloured}
\end{subfigure}%
\begin{subfigure}
  \centering
  \includegraphics[width=0.55\textwidth]{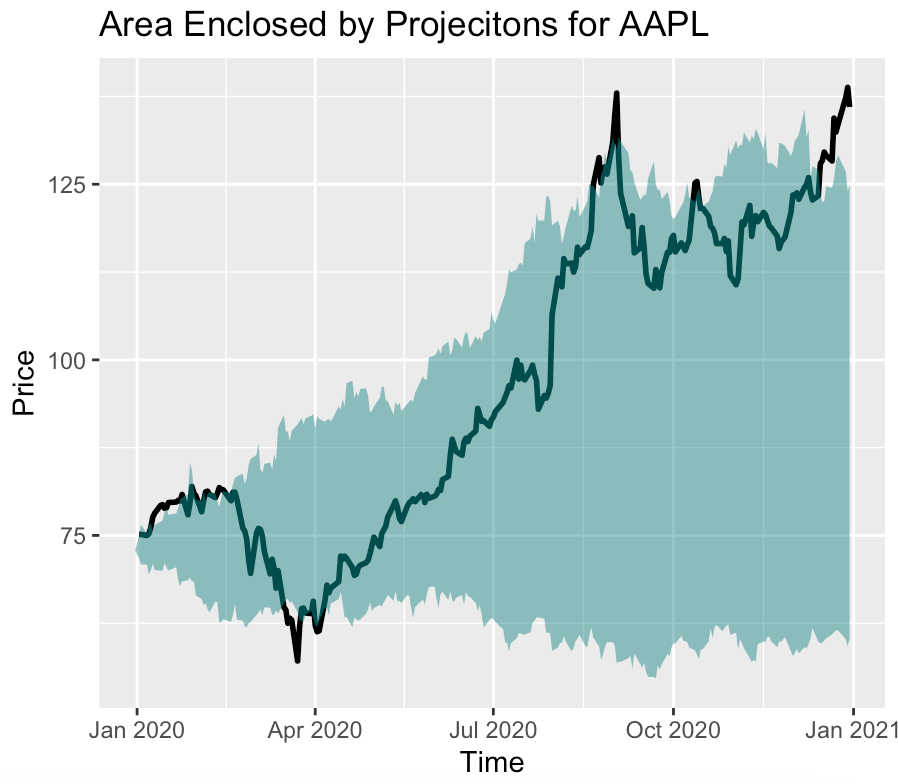}
\end{subfigure}
\caption{Plots of (Left) Apple Stock Price (Black) from January 2020 to December 31 2020 and projections (coloured) and (Right) historical Apple stock price from January 2020 to January 2021 and the area between the projections assuming the law in equation \eqref{eq:law Apple} with $\Gamma = 0.0008316271$ and $\Sigma = 0.01648899$.}
\label{fig:Apple}
\end{figure}

\begin{figure}[H]
    \centering
    \includegraphics[width = 0.7\textwidth]{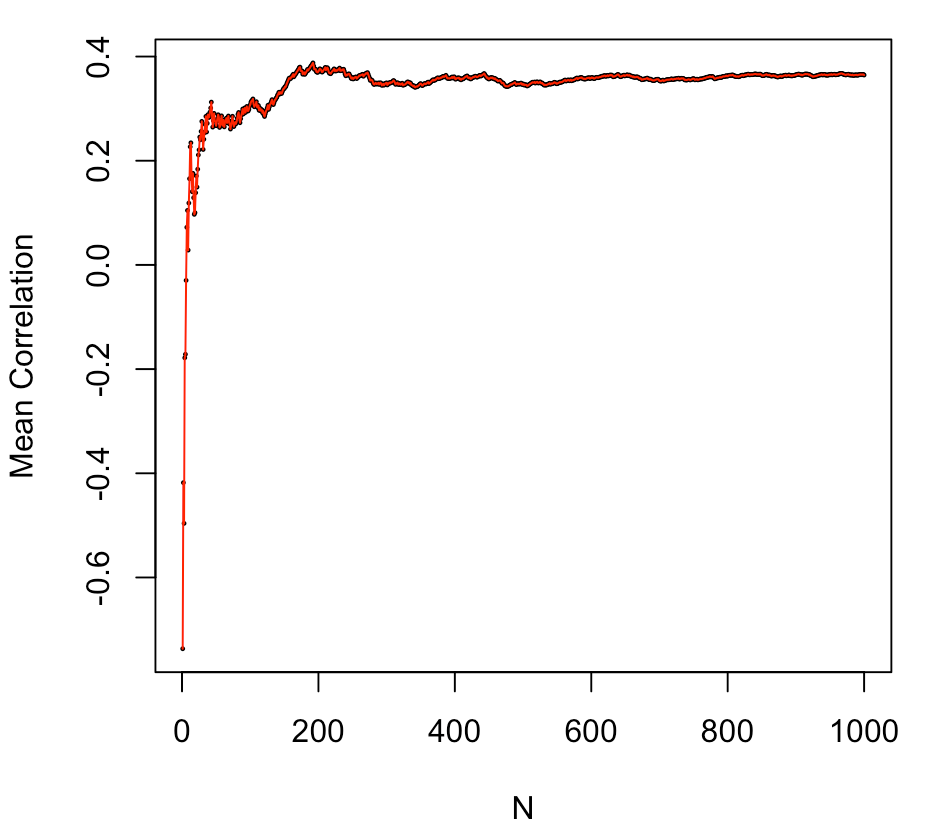}
    \caption{Plot of mean correlation $\mathfrak{Cor}^{N}$ for values of N in $\{1,2,3,...., 1000\}$. The plot seems to converge to a value of approximately $0.36$}
    \label{fig: mean correlation}
\end{figure}

Now, for some numerical results to buttress the theory, we consider the case where the stock price process $\mathcal Y_t$ from some initial time $t = 0$ to some final time $t = T$ follows the law 

\begin{equation}
    \displaystyle \log \mathcal Y_{t} - \log \mathcal Y_{0} 
     = \int_{0}^{t}\Gamma dt + \int_{0}^{t}\Sigma d \mathfrak B_t
     \label{eq:law Apple}
\end{equation} for $t \in [0,T]$, where $\gamma(t) = \Gamma$ and $\sigma(\mathcal Y_t, t) = \Sigma$ are constants. The above formula is well defined as constant functions are members of the space $\mathfrak L(0,T)$ for which the Itô integral is defined. The above can be solved analytically, but a numerical treatment will be explored within the context of Apple's stock price.\\
\\In the following example, the model will be tested against historical data from January 1 2020 to December 31 2020. $\Gamma$ and $\Sigma$ will be computed using historical daily log-returns from 11 October 2007. By daily log-returns at a given date, we mean the natural logarithm of the ratio of the price at said day by the price at the previous date. \\
\\First, we partition the period from 1 January 2020 to 31 December 2020 into $\mathfrak T = \{T_i\}_{i \equiv i^{th} \text{trading day in the year}}$ starting from $i = 0$ and we call the collection of log-returns $\mathfrak{logR} = \{\log R_{t_i}\}_{t_i \in \mathfrak T\symbol{92}\{T_0\}}$ where $R = \frac{X_i}{X_{i-1}}$ and $X_i$ is the price of Apple's stock at some $t_i \in \mathfrak T \symbol{92}\{T_0\}$. We use the above to compute $\Gamma$ and $\Sigma$:

\begin{gather*}
\centering
 \left\{
  \begin{array}{@{}ll@{}}
    \Gamma \\
    \Sigma
  \end{array}\right.
  = 
  \left\{
  \begin{array}{@{}ll@{}}
    \overline{\mathfrak{LogR}} - \frac{\text{Var}(\mathfrak{LogR})^{2}}{2}\ \\
    \text{Var}(\mathfrak{LogR})
  \end{array}\right.
\end{gather*} 
where $\overline{\mathfrak{logR}}$ is the (sample) mean and $\text{Var}(\mathfrak{logR})$ the (sample) variance of $\mathfrak{logR}$ respectively.\\
\\We are now in a position to consider discrete approximation to equation \ref{eq: sde}. As per section \ref{Application of SDE}, we consider the discrete stock price process $\mathcal Y_{t}|_{\mathfrak T} : \mathfrak T \times \Omega \rightarrow \mathbb R$ given by: 
\begin{gather*}
    \mathcal Y_{t_n}|_{\mathfrak T}(\omega) \\
    = H_{t_0} \times \exp{\left(\sum_{k=0}^{n-1}\gamma(t_{k-1})(\omega)\cdot (t_k-t_{k-1})\right)}\\ \times \exp{\left( \sum_{k=0}^{n-1} \sigma(t_{k-1}, \mathcal Y_{t_{k-1}}(\omega))\cdot(\mathfrak B_{t_k}(\omega)-\mathfrak B_{t_{k-1}}(\omega)) \right)}
\end{gather*} where $\mathcal Y_{t_{0}}|_{\mathfrak T} = H_{t_0}$, $t_n \in \mathfrak T$ and $\omega \in \Omega$.

\noindent
\\It is clear from the sub-figure on the right of figure \ref{fig:Apple}, the simulated paths capture most of the historical time series of Apple's stock price which is another indicator that the model is qualitatively speaking, a good approximation. However, only ten projections were performed, justifiably casting doubt on the statistical significance of the above result.\\
\\To try and quantify the above intuition, we will try and compute numerically the expected value, a central tendency indicator, $\mathbb{E} \left[\text{Cor}(\mathcal Y_t|_{\mathfrak T},H_t)\right]$ of the correlation between $\mathcal Y_{t}|_{\mathfrak T}$ and $H_t$.\\
\\To achieve this, we consider $N$ independent and identically distributed copies of the discrete process $\mathcal Y_{t}|_{\mathfrak T}$, $\mathcal Y_{t}^{n}|_{\mathfrak T}$ where $n \in [1,N]\cap\mathbb{N}$; intuitively, they correspond to $N$ distinct projections. The correlation between the historical data and the  $n^{\text{th}}$ projection is denoted by
\begin{equation*}
    \text{Cor}(\mathcal Y_t^{n}|_{\mathfrak T}, H_t)
\end{equation*}
\noindent
with cumulative mean
\begin{equation*}
    \displaystyle \mathfrak{Cor}^{N} = \frac{\sum_{n=1}^{N}\text{Cor}(\mathcal Y_t^{n}, H_t)}{N}
\end{equation*} taking $N \rightarrow \infty$, we obtain figure \ref{fig: mean correlation}.\\

\noindent
\\According to basic probability theory, $\mathfrak{Cor}^N$ converge in the sense of probability to the expected value of the correlation between the discretised path and the historical trial data $\mathbb{E} \left[\text{Cor}(\mathcal{Y}_t|_{\mathfrak T},H_t)\right]$ \cite{blitzstein2019introduction}. This means that considering progressively larger values of N (that is taking $N\rightarrow \infty$), $\mathfrak{Cor}^N$ should approach a constant value (see figure \ref{fig: mean correlation}).

\noindent
Figure \ref{fig: mean correlation} suggests that $\mathbb{E} \left[\text{Cor}(\mathcal \mathcal{Y}_t|_{\mathfrak T},H_t)\right] \approx \mathfrak{Cor}^{N}$ tends to a constant value of $0.36$ for $N$ large. This suggests, in a rather heuristic way, that the model 'captures' 36\% of the variability in stock prices. Granted, further investigation is warranted over different stocks and more involved models could of course be considered.
\section{Conclusion}

Through the course of the above, it has become palpable how Brownian Motion encapsulates and can be used to model a crucial part of the uncertainty inherent in financial markets paving way for a stochastic analysis, culminating in the development of the Itô integral.

\bibliographystyle{plain} 
\bibliography{refs} 

\begin{thebibliography}{1}

\bibitem{2006}
{\em Introduction to Stochastic Integration}.
\newblock Springer-Verlag, 2006.

\bibitem{bachelier1900theorie}
Louis Bachelier.
\newblock Th{\'e}orie de la sp{\'e}culation.
\newblock In {\em Annales scientifiques de l'{\'E}cole normale sup{\'e}rieure},
  volume~17, pages 21--86, 1900.

\bibitem{blitzstein2019introduction}
Joseph~K Blitzstein and Jessica Hwang.
\newblock {\em Introduction to probability}.
\newblock Chapman and Hall/CRC, 2019.

\bibitem{bremaud2020probability}
Pierre Br{\'e}maud.
\newblock {\em Probability Theory and Stochastic Processes}.
\newblock Springer Nature, 2020.

\bibitem{brown1828xxvii}
Robert Brown.
\newblock Xxvii. a brief account of microscopical observations made in the
  months of june, july and august 1827, on the particles contained in the
  pollen of plants; and on the general existence of active molecules in organic
  and inorganic bodies.
\newblock {\em The philosophical magazine}, 4(21):161--173, 1828.

\bibitem{fernholz2002stochastic}
E~Robert Fernholz.
\newblock Stochastic portfolio theory.
\newblock In {\em Stochastic portfolio theory}, pages 1--24. Springer, 2002.

\bibitem{kolmogorov2018foundations}
Andrei~Nikolaevich Kolmogorov and Albert~T Bharucha-Reid.
\newblock {\em Foundations of the theory of probability: Second English
  Edition}.
\newblock Courier Dover Publications, 2018.

\bibitem{oksendal2003stochastic}
Bernt {\O}ksendal.
\newblock Stochastic differential equations.
\newblock In {\em Stochastic differential equations}, pages 65--84. Springer,
  2003.

\bibitem{wiener1923differential}
Norbert Wiener.
\newblock Differential-space.
\newblock {\em Journal of Mathematics and Physics}, 2(1-4):131--174, 1923.

\end{thebibliography}
\end{document}